\documentclass[10pt,twocolumn]{article}

\usepackage[utf8]{inputenc}
\usepackage{amsmath}
\usepackage{amsthm}
\usepackage{amssymb}
\usepackage{titlesec}
\usepackage{cite}
\usepackage{booktabs}
\usepackage{graphicx}
\usepackage[colorlinks=false]{hyperref}
\usepackage[format=plain,labelfont=it]{caption}
\usepackage[left=1.5cm,right=1.5cm,top=2cm,bottom=2cm]{geometry}
\usepackage{color}

\titleformat{\section}{\centering\normalfont\scshape}{\Roman{section}.}{5pt}{}
\titleformat{\subsection}{\normalfont\it}{\Alph{subsection}.}{5pt}{}
\titleformat{\subsubsection}{\normalfont\it}{\hspace{4mm}\arabic{subsubsection})}{5pt}{}

\newcommand\infoFootnote[1]{%
  \begingroup
  \renewcommand\thefootnote{}\footnote{#1}%
  \addtocounter{footnote}{-1}%
  \endgroup}

\newtheorem{thm}{Theorem}
\newtheorem{cor}[thm]{Corollary}
\newtheorem{lem}[thm]{Lemma}
\newtheorem{prop}[thm]{Proposition}
\newtheorem{assum}{Assumption}

\DeclareMathOperator*{\argmin}{\arg\min}

\newcommand{\R}{\mathbb{R}} 

\newcommand{\ab}{\boldsymbol{a}}

\newcommand{\ub}{\boldsymbol{u}}

\newcommand{\xb}{\boldsymbol{x}}
\newcommand{\yb}{\boldsymbol{y}}
\newcommand{\xib}{\boldsymbol{\xi}}
\newcommand{\zerob}{\boldsymbol{0}}

\newcommand{\Ib}{\boldsymbol{I}}
\newcommand{\Kb}{\boldsymbol{K}}
\newcommand{\Lb}{\boldsymbol{L}}

\newcommand{\Qb}{\boldsymbol{Q}}

\newcommand{\Ub}{\boldsymbol{U}}

\newcommand{\Wb}{\boldsymbol{W}}

\newcommand{\Yb}{\boldsymbol{Y}}
\newcommand{\Zb}{\boldsymbol{Z}}
\newcommand{\Pib}{\boldsymbol{\Pi}}

\newcommand{\gammab}{\boldsymbol{\gamma}}

\newcommand{\ybs}{\mathbf{y}}

\newcommand{\ubs}{\mathbf{u}}

\newcommand{\Dbc}{\boldsymbol{\mathcal{D}}}

\newcommand{\Qbc}{\boldsymbol{\mathcal{Q}}}
\newcommand{\Rbc}{\boldsymbol{\mathcal{R}}}

\newcommand{\Uc}{\mathcal{U}}

\newcommand{\Yc}{\mathcal{Y}}

\renewcommand{\boldsymbol}[1]{#1}
\renewcommand{\mathbf}[1]{\mathrm{#1}}

\theoremstyle{definition}

\title{\vspace{-2mm}\bf Towards a unifying framework for data-driven predictive control with quadratic regularization}
\author{Manuel Kl\"adtke and Moritz Schulze Darup\vspace{2mm}}
\date{}

\begin{document}

\maketitle

\textbf{\textit{Abstract}.} {\bf Data-driven predictive control (DPC) has recently gained popularity as an alternative to model predictive control (MPC). Amidst the surge in proposed DPC frameworks, upon closer inspection, many of these frameworks are more closely related (or perhaps even equivalent) to each other than it may first appear. We argue for a more formal characterization of these relationships so that results can be freely transferred from one framework to another, rather than being uniquely attributed to a particular framework. We demonstrate this idea by examining the connection between $\gammab$-DDPC and the original DeePC formulation.}
\infoFootnote{M. Kl\"adtke and M. Schulze Darup are with the \href{https://rcs.mb.tu-dortmund.de/}{Control and~Cyberphysical Systems Group}, Faculty of Mechanical Engineering, TU Dortmund University, Germany. E-mails:  \href{mailto:manuel.klaedtke@tu-dortmund.de}{\{manuel.klaedtke,  moritz.schulzedarup\}@tu-dortmund.de}. \vspace{0.5mm}}
\infoFootnote{\hspace{-1.5mm}$^\ast$This paper is a \textbf{preprint} of a contribution to the 26th International Symposium on Mathematical Theory of Networks and Systems (MTNS).}

\section{Introduction}

Data-driven predictive control (DPC) is an increasingly popular control approach that utilizes linear combinations of collected trajectory data to make predictions instead of relying on a system model (see, e.g., \cite{Coulson2019DeePC, Berberich2020}). Coming with its popularity, there is a surge in different DPC frameworks (e.g., \cite{Breschi2022new,Lazar2023GeneralizedDP}) proposing modifications to the original DeePC formulation \cite{Coulson2019DeePC}. While each of these frameworks allows for interesting results, and some relationships between them are already established, we are convinced that many connections remain yet undiscovered. Rather than arguing for or against a particular framework, we advocate for a formal characterization of such relationships in order to facilitate the transfer of results and to unify them. In this note, we demonstrate this concept by examining the connection between $\gammab$-DDPC \cite{Breschi2022new} and DeePC \cite{Coulson2019DeePC} with quadratic regularization.

The note is organized as follows. In Section~\ref{sec:fundamentals}, we summarize fundamentals of DPC and introduce the DeePC as well as the $\gammab$-DDPC formulations. In Section~\ref{sec:revisitingReg}, we revisit well established quadratic regularization strategies for DeePC and give new interpretations to their associated costs. In Section~\ref{sec:gamma-DDPC}, we establish equivalences between proposed quadratic regularization strategies for the DeePC and $\gammab$-DDPC frameworks. Finally, we preview future opportunities for unification of DPC frameworks in Section~\ref{sec:conclusions}.

\section{Fundamentals of DPC} \label{sec:fundamentals} 

Instead of utilizing a discrete-time state-space model with input $\ub\in\R^m$, state $\xb\in\R^n$, and output $\yb\in\R^p$ as in traditional model predictive control (MPC), predictions in DPC are realized based on previously collected trajectory data $(\ubs^{(1)}, \ybs^{(1)}), \hdots, (\ubs^{(\ell)}, \ybs^{(\ell)})$ via linear combinations
$$
    \begin{pmatrix}
        \ubs_\text{pred}\\
        \ybs_\text{pred}
    \end{pmatrix}
    =
    \begin{pmatrix}
        \ubs^{(1)}\\
        \ybs^{(1)}
    \end{pmatrix} a_1 + \hdots + 
    \begin{pmatrix}
        \ubs^{(\ell)}\\
        \ybs^{(\ell)}
    \end{pmatrix} a_\ell
    = 
    \Dbc \ab.
$$
Here, the dimensions of the data matrix $\Dbc\in\R^{L(m+p)\times\ell}$ and generator vector $\ab\in\R^\ell$ are specified by the length~$L$ of recorded (and predicted) trajectories and the number~$\ell$ of data trajectories used for predictions. To include the current initial condition of the system as a starting point for predicted trajectories, the I/O-sequence is typically partitioned into a past section $(\ubs_p, \ybs_p)$ and a future section $(\ubs_f, \ybs_f)$ with $N_p$ respectively $N_f$ time-steps yielding
$$
    \begin{pmatrix}
        \ubs_p\\
        \ubs_f
    \end{pmatrix} = \ubs_\text{pred} 
    = 
    \begin{pmatrix}
        \Ub_p\\
        \Ub_f
    \end{pmatrix} \ab, \quad 
    \begin{pmatrix}
        \ybs_p\\
        \ybs_f
    \end{pmatrix} = \ybs_\text{pred} 
    = 
    \begin{pmatrix}
        \Yb_p\\
        \Yb_f
    \end{pmatrix} \ab. 
$$
The past section of a predicted trajectory is then forced to match the I/O-data $\xib$ recorded in the most recent $N_p$ time-steps during closed-loop operation, i.e., the constraints 
$$
    \xib= 
    \begin{pmatrix}
        \ubs_p \\ \ybs_p
    \end{pmatrix}
    =
    \begin{pmatrix}
        \Ub_p \\ \Yb_p
    \end{pmatrix}\ab
    = \Wb_p \ab 
$$
force any predicted trajectory to start with the most recently witnessed behavior of the system. 
Furthermore, we define the shorthand notation $\Zb := \begin{pmatrix}
        \Wb_p^\top & \Ub_f^\top
    \end{pmatrix}\!^\top.$

\subsection{Fundamentals of DeePC and underlying assumption}

The optimal control problem (OCP) that is solved for DeePC in every time step can be stated as 
\begin{subequations}
\label{eq:DeePC}
\begin{align}
\min_{\ubs_f,\ybs_f,\ab} 
 J(\xib, \ubs_f, &\ybs_f) + h(\ab) \label{eq:DPCcost} \\
\text{s.t.} \quad \quad  \begin{pmatrix}
    \xib \\ \ubs_f \\ \ybs_f 
\end{pmatrix} &= \begin{pmatrix}
    \Wb_p \\ \Ub_f \\ \Yb_f 
\end{pmatrix}\ab, \label{eq:DPCeqConstr}\\
\left(\ubs_f, \ybs_f \right) &\in \Uc \times \Yc.  \label{eq:DPCsetConstr}
\end{align}
\end{subequations}
with control objective $J(\xib, \ubs_f, \ybs_f)$, regularization $h(\ab)$, and input-output constraints $\Uc \times \Yc$. Conditions for equivalence of \eqref{eq:DeePC} and MPC are well established for some cases, where the baseline equivalence with classical linear MPC requires having exact and persistently exciting data \cite{WILLEMS2005}, the underlying system being LTI, and setting the regularization to $h(\ab) = 0$ \cite{Coulson2019DeePC}. Here, we only make the following assumption.
\begin{assum}\label{assum:fullRank}
    The data matrix $\Dbc$ has full row rank.
\end{assum}
This assumption aligns not only with the case of an LTI system with noisy output measurements as assumed in the $\gammab$-DDPC framework \cite[Lem.~3]{Breschi2022new} but also encompasses many more. Crucially, this highlights the universal character of the following analysis, which depends more on the structure of the OCP rather than on the specific nature of data in $\Dbc$. 
For a more detailed discussion, see  \cite[Sect.~III]{KLAEDTKE2023}.
Crucially, Assumption~\ref{assum:fullRank} allows for the following characterization of regularization costs in terms of actual system trajectories, which we will use in Section~\ref{sec:revisitingReg}.
\begin{lem}[\cite{KLAEDTKE2023}]\label{lem:priceTag}
    Under Assumption~\ref{assum:fullRank}, the DeePC problem \eqref{eq:DeePC} is equivalent to   
    \begin{equation}\label{eq:regDPCouter}
    \min_{\ubs_f,\ybs_f}
      J(\xib, \ubs_f, \ybs_f) + h^\ast(\xib, \ubs_f, \ybs_f) \quad \text{s.t.}\quad  \eqref{eq:DPCsetConstr}
    \end{equation}
    with unique
    \begin{equation}\label{eq:regDPCinner}
    h^\ast(\xib, \ubs_f, \ybs_f) := \min_{\ab}\:\:   h(\ab) \quad 
    \text{s.t.}\quad \eqref{eq:DPCeqConstr}. 
    \end{equation}
\end{lem}

\subsection{From DeePC to $\gammab$-DDPC}

The $\gammab$-DDPC framework introduced in \cite{Breschi2022new} is based on an LQ decomposition of the data matrices and subsequent coordinate transformation of the optimization variable $\ab$, which replaces \eqref{eq:DPCeqConstr} as follows 
$$
\begin{pmatrix}
        \Wb_p \\
        \Ub_f \\
        \Yb_f
    \end{pmatrix} \!
    =  \!
    \begin{pmatrix}
        \Lb_{11} & \zerob & \zerob & \zerob \\
        \Lb_{21} & \Lb_{22} & \zerob & \zerob \\
        \Lb_{31} & \Lb_{32} & \Lb_{33} & \zerob 
    \end{pmatrix}
    \begin{pmatrix}
        \Qb_1 \\
        \Qb_2 \\
        \Qb_3 \\
        \Qb_4 
    \end{pmatrix}, 
    \; \begin{pmatrix}
        \gammab_1 \\ 
        \gammab_2 \\
        \gammab_3 \\
        \gammab_4 \\
    \end{pmatrix} \! = \! \begin{pmatrix}
        \Qb_1 \\
        \Qb_2 \\
        \Qb_3 \\
        \Qb_4 
    \end{pmatrix} \ab. 
$$
Here, the diagonal blocks $\Lb_{ii}$ for $i \in \{1, 2, 3\}$ are non-singular and the matrices $\Qb_i$ have orthonormal rows, i.e., $\Qb_i \Qb_i^\top = \Ib$ and $\Qb_i \Qb_j^\top = \zerob$ for $i \neq j$. Note that, technically, the data matrix is normalized w.r.t. $\sqrt{\ell}$ in the $\gammab$-DDPC framework. However, the same effect can be achieved by rescaling regularization weights with $\ell$. Also, $\Qb_4$ and $\gammab_4$ are typically omitted, which corresponds to constraining $\gammab_4 = \zerob$. Furthermore, $\gammab_1=\Lb_{11}^{-1}\xib$ is uniquely determined by the current state $\xib$, which leaves $\gammab_2\in \R^{m N_f}$ and $\gammab_3 \in \R^{p N_f}$ as optimization variables that are regularized via a term $\tilde h(\gammab)$ replacing $h(\ab)$ in \eqref{eq:DPCcost}. In the subsequent works \cite{BRESCHI2023uncertaintyAware, Breschi2023regularizationImpact}, the authors examine regularization strategies based on a mix of quadratic regularization for $\gammab_2, \gammab_3$, or constraining $\gammab_3 = \zerob$. 
\subsection{A close relative: Subspace predictive control (SPC)}
An indirect data-driven alternative to DPC is given by subspace predictive control (SPC, \cite{FAVOREEL1999}), where the constraint \eqref{eq:DPCeqConstr} is replaced by the linear multi-step predictor
$$
    \ybs_f = \hat\ybs_{\text{SPC}}(\xib,\ubs_f) = \Kb_{\text{SPC}} \begin{pmatrix}
        \xib \\ \ubs_f
    \end{pmatrix},
$$
where, in its most basic form, 
\begin{equation}
    \Kb_{\text{SPC}} := \argmin_\Kb \left\|\Yb_f-\Kb \Zb \right\|_F^2 = \Yb_f \Zb^+ \label{eq:K_SPC} 
\end{equation}
is the solution to a least squares problem with the Frobenius norm $\|\cdot\|_F$. While we do not actively analyze SPC in this note, it has an intrinsic connection to DPC with quadratic regularization as established in \cite{Dorfler2021}, which will be discussed in the following sections.

\section{Revisiting quadratic regularization for DeePC}\label{sec:revisitingReg}

In this section, we analyze the regularization effect of the two popular choices $h(\ab) = \lambda_a \|\ab\|_2^2$ and $h(\ab) = \lambda_a \|\Pi_\perp \ab\|_2^2$ with projection matrices 
$$
    \Pib := \Zb^+\Zb \quad \text{and} \quad  \Pi_\perp := \Ib - \Pib . 
$$ 
While the characterization in terms of Lemma~\ref{lem:priceTag} is already given as an intermediate result in \cite[Sect.~III.A-B]{KLAEDTKE2023} for both choices, we will revisit them to provide interpretations of the corresponding cost terms. Furthermore, while  these expressions follow directly from block matrix inversion formulas, we will omit their derivation due to space restrictions. 

\subsection{Standard quadratic regularization} \label{sec:standard2Norm}

The choice $h(\ab) = \lambda_a \|\ab\|_2^2$ was initially proposed in \cite{Coulson2019DeePC} as a heuristic, with interpretations for its good performance given shortly after (e.g., \cite{Coulson2019RegularizedDeePC, Dorfler2021}). In terms of Lemma~\ref{lem:priceTag}, its effects can be characterized as follows.
\begin{prop}\label{prop:2NormCost}
    The effect of $h(\ab) = \lambda_a \|\ab\|_2^2$ in the sense of Lemma~\ref{lem:priceTag} is given by 
    \begin{subequations}\label{eq:2NormCost}
        \begin{align}
        h^\ast(\xib, \ubs_f, \ybs_f) &= \lambda_a \|\ybs_f-\hat\ybs_\text{SPC}(\xib, \ubs_f)\|_{\Qbc_\text{reg}}^2 \label{eq:2NormCostTerm1}\\
    &\qquad + \lambda_a \|\ubs_f-\Ub_f \Wb_p^+\xib\|_{\Rbc_\text{reg}}^2 \label{eq:2NormCostTerm2}\\
    & \qquad+ \lambda_a\|\xib\|_{\left(\Wb_p\Wb_p^\top\right)^{-1}}^2 \label{eq:2NormCostTerm3}
    \end{align}
    \end{subequations}
    with weighing matrices $\Qbc_\text{reg} :=\left(\Yb_f \left(\Ib - \Pib\right) \Yb_f^\top\right)^{-1}$ and 
    $$
    \Rbc_\text{reg} :=\left(\Ub_f\left(\Ib-\Wb_p^+\Wb_p\right)\Ub_f^\top\right)^{-1}.
    $$
\end{prop}
While the last cost term \eqref{eq:2NormCostTerm3} is irrelevant, since $\xib$ is not an optimization variable, the other two terms offer a lot of insightful interpretations. The first term  \eqref{eq:2NormCostTerm1} regularizes predicted output trajectories $\ybs_f$ towards the SPC predictor $\hat\ybs_\text{SPC}(\xib, \ubs_f)$.
 For high $\lambda_a$, this term can even be seen as soft constraint version of the SPC equality constraint $\ybs_f = \hat\ybs_\text{SPC}(\xib, \ubs_f)$ that has been lifted to the cost function. 
One effect of this soft constraint is the unintuitive interaction of DPC with output constraints (e.g., $\ybs_f \in \Yc$) observed in \cite[Sect.~III.C]{KLAEDTKE2023}. The weighing matrix $\Qbc_\text{reg}$ is best understood by treating $\Yb_f \Pib_\perp = \Yb_f \left(\Ib-\Pib\right) ={\Yb_f-\Kb_\text{SPC}\Zb }= \Yb_f-\Yb_{f\text{,SPC}}  = \Delta \Yb_f$ itself as a data matrix with columns containing data of the prediction error $\Delta \ybs_f= \ybs_f - \hat\ybs_\text{SPC}(\xib, \ubs_f)$ w.r.t. the least squares estimate given by $\hat\ybs_\text{SPC}(\xib, \ubs_f)$. Note that we can alternatively express the weighing matrix as 
$$
    \Qbc_{reg} = \left(\Delta\Yb_f \Delta\Yb_f^\top \right)^{-1} 
$$
because $\Pib_\perp$ is an orthogonal projection matrix. 
Since $\Delta\Yb_f$ is a data matrix of the SPC prediction error, the matrix $\Delta\Yb_f \Delta\Yb_f^\top $ is a scaled empirical second moment matrix of this data. Therefore, while \eqref{eq:2NormCostTerm1} generally regularizes output predictions towards the SPC predictor, it does so more (less) harshly in directions, where the SPC predictor is believed to be more (less) accurate, based on the available data. Note that the weighing matrix is neither normalized w.r.t. the number $\ell$ of data trajectories nor the general magnitude of control objective costs, which should be kept in mind when tuning $\lambda_a$.

Similar to the output $\ybs_f$ being regularized towards the least squares estimate $\hat\ybs_\text{SPC}(\xib, \ubs_f)$ in \eqref{eq:2NormCostTerm1}, the input $\ubs_f$ is regularized towards the least squares estimate $\Ub_f \Wb_p^+ \xib$ in \eqref{eq:2NormCostTerm2}. 
This term may be interpreted as regularizing the chosen input towards the best-explored region of the state-input-space, i.e., where the most confident predictions can be made based on the available data. This interpretation is supported by 
$$
    \Rbc_{reg} = \left(\Delta\Ub_f \Delta\Ub_f^\top \right)^{-1},
$$
again, being the inverse of a scaled empirical second moment matrix for the ``input prediction error'' data ${\Delta \Ub_f = \Ub_f - \Ub_f\Wb_p^+ \Wb_p}$. 
This cost term can indeed be utilized to increase DPC performance, as analyzed in \cite{BRESCHI2023uncertaintyAware, Breschi2023regularizationImpact}, and we show how to translate these results to DeePC in Section~\ref{sec:gamma-DDPC}. 
However, this part of the cost also renders the closed-loop control cautious and exploration-averse, so a large weight on this term might be unsuited for DPC schemes, where the trajectory data is continuously updated during closed-loop operation. Furthermore, if the data trajectories were recorded in closed-loop, the term $\Ub_f \Wb_p^+$ implicitly estimates a linear controller that matches this closed-loop collection phase. Note that this still fits the ``cautious input regularization'' interpretation, since the predicted optimal inputs are pushed towards the best-explored area of the state-input-space, i.e., the subspace associated with applying the estimated linear controller.
Finally, having only one tuning parameter $\lambda_a$ affecting these two fundamentally different regularization terms may lead to problems. For example, one might be inclined to enforce rather strict agreement with the SPC predictor via \eqref{eq:2NormCostTerm1}, while only wishing for slight regularization of the input via \eqref{eq:2NormCostTerm2}. This is also why the regularization $h(\ab) = \lambda_a \|\ab\|_2^2$ is typically outperformed by $h(\ab) = \lambda_a \|\Pi_\perp \ab\|_2^2$ for $\lambda_a \to \infty$ (see, e.g., \cite[Fig.~2]{Dorfler2021}). However, a mix of the two regularizations may offer the best of both worlds, which is coincidentally captured by the regularization strategies proposed for $\gammab$-DDPC \cite{BRESCHI2023uncertaintyAware} analyzed in Section~\ref{sec:gamma-DDPC}.

\subsection{Projection-based quadratic regularization} \label{sec:proj2Norm}

The projection-based choice $h(\ab) = \lambda_a \|\Pi_\perp \ab\|_2^2$ was proposed in \cite{Dorfler2021} to remove the inconsistency of the norm-based regularization $h(\ab) = \lambda_a \| \ab\|_2^2$ w.r.t. the least-square estimate given by the SPC predictor $\hat\ybs_\text{SPC}(\xib, \ubs_f)$. Indeed, as specified in the following proposition,  the projection leaves the first term \eqref{eq:2NormCostTerm1} intact, while removing the other two terms. 
\begin{prop}
    The effect of $h(\ab) = \lambda_a \|\Pi_\perp\ab\|_2^2$ in the sense of Lemma~\ref{lem:priceTag} is given by 
    \begin{equation}
        h^\ast(\xib, \ubs_f, \ybs_f) = \lambda_a \|\ybs_f-\hat\ybs_\text{SPC}(\xib, \ubs_f)\|_{\Qbc_\text{reg}}^2 \label{eq:proj-2NormCost} 
    \end{equation}
    with $\Qbc_\text{reg}$ as in Proposition~\ref{prop:2NormCost}.
\end{prop}
The interpretation of the cost term in~\eqref{eq:proj-2NormCost} carries over one to one from the comments made below Proposition~\ref{prop:2NormCost}.

\section{Relation to $\gammab$-DDPC} \label{sec:gamma-DDPC}

While the LQ decomposition is a useful preprocessing step, it only amounts to a coordinate transformation for the optimization variable and thus does not constitute a fundamentally new approach to DPC. We support this claim by first showing how the regularization strategies analyzed in Section~\ref{sec:standard2Norm}~and~\ref{sec:proj2Norm} can be reproduced in the $\gammab$-DDPC framework and second, conversely, how the regularization strategies in \cite{BRESCHI2023uncertaintyAware, Breschi2023regularizationImpact} can be reproduced in the original DeePC setting. Note that ``equivalence of OCPs'' in the following propositions means that they yield the same optimal predicted I/O trajectories $(\ubs_f^\ast, \ybs_f^\ast)$ for every state $\xib$ (which, crucially, also implies equivalent closed-loop behavior of the control schemes), but not necessarily the same optimal cost.

\begin{prop}\label{prop:DDPCequivalence1}
    DeePC with regularization $h(\ab) = \lambda_a \|\ab\|_2^2$ is equivalent to $\gammab$-DDPC with $\tilde h(\gammab) = \lambda_a \|\gammab_2\|_2^2 + \lambda_a \|\gammab_3\|_2^2$.
\end{prop}
\begin{proof}
First, note that $\Qb^\top = \begin{pmatrix}
    \Qb_1^\top & \Qb_2^\top & \Qb_3^\top & \Qb_4^\top 
\end{pmatrix}$
is an orthonormal basis of $\R^\ell$ and therefore $\Qb^\top \Qb = \Ib$. Hence, we can equivalently express 
\begin{equation}
    \|\ab\|_2^2 = \ab^\top \Ib \ab =\ab^\top \Qb^\top \Qb \ab = \sum_{i = 1}^4 \|\gammab_i\|_2^2 \label{eq:2norm2gamma} 
\end{equation}
However, $\gammab_1$ is not an optimization variable, since it is uniquely determined by $\xib$ and therefore irrelevant to the optimal predicted I/O trajectories $(\ubs_f^\ast, \ybs_f^\ast)$. Similarly, $\gammab_4$ does not affect any $(\xib, \ubs_f, \ybs_f)$, since the corresponding $\Qb_4$ only characterizes the null space of the data matrices. More specifically, penalizing $\|\gammab_4\|_2^2$ always yields the optimizer $\gammab_4^\ast = \zerob$, which is already inherently enforced in $\gammab$-DDPC by dropping $\Qb_4$ and $\gammab_4$.  
\end{proof}

For clarity of the upcoming proof, we introduce the shorthand notation  
$$\Qb_{1:2} := \begin{pmatrix}
    \Qb_1 \\ \Qb_2 
\end{pmatrix} \quad \text{and} \quad 
\Lb_{1:2} :=
\begin{pmatrix}
 \Lb_{11} & \zerob & \\
        \Lb_{21} & \Lb_{22}         
\end{pmatrix}$$
and point out that $\Zb = \Lb_{1:2}\Qb_{1:2}$ and ${\Qb_{1:2} \Qb_{1:2}^\top = \Ib}$. Furthermore, $\Lb_{1:2}$ is non-singular.

\begin{prop}\label{prop:DDPCequivalence2}
    DeePC with regularization $h(\ab) = \lambda_a \|\Pib_\perp\ab\|_2^2$ is equivalent to $\gammab$-DDPC with $\tilde h(\gammab) = \lambda_a \|\gammab_3\|_2^2$.
\end{prop}
\begin{proof}
    First, note that 
    \begin{align}
        \Pib &= \Zb^+\Zb = \Zb^\top \left(\Zb \Zb^\top\right)^{-1}\Zb  \label{eq:PiViaGamma}\\
        &= \Qb_{1:2}^\top \Lb_{1:2}^\top \left(\Lb_{1:2} \Qb_{1:2} \Qb_{1:2}^\top \Lb_{1:2}^\top\right)^{-1} \Lb_{1:2} \Qb_{1:2} = \Qb_{1:2}^\top \Qb_{1:2} \nonumber
    \end{align}
    and hence
    $$
        \Pib_\perp = \Ib-\Pib = \Qb^\top \Qb - \Qb_{1:2}^\top \Qb_{1:2} = \Qb_3^\top\Qb_3 +  \Qb_4^\top\Qb_4.
    $$
    We can therefore equivalently express 
    \begin{align}
    \|\Pib_\perp \ab\|_2^2 = \ab^\top\!\! \left(\Qb_3^\top\Qb_3 +  \Qb_4^\top\Qb_4\right) \ab 
    =  \|\gammab_3\|_2^2 + \|\gammab_4\|_2^2. \label{eq:2normProj2gamma}
\end{align}
    The same considerations as in Proposition~\ref{prop:DDPCequivalence1} apply regarding $\gammab_4$, hence the claim follows.
\end{proof}  
 Conversely, the following two propositions concern reformulations of the two regularization strategies proposed in \cite{BRESCHI2023uncertaintyAware}.
 \begin{prop}\label{prop:DDPCstrategy1}
     $\gammab$-DDPC with ${\tilde  h(\gammab) = \lambda_2 \|\gammab_2\|_2^2 + \lambda_3 \|\gammab_3\|_2^2}$ is equivalent~to~DeePC~with $h(\ab) = \!\lambda_2 \|\ab\|_2^2 + {(\lambda_3\!-\!\lambda_2)} \|\Pib_\perp\ab\|_2^2$.
 \end{prop}
\begin{proof}
    By combining \eqref{eq:2norm2gamma} and \eqref{eq:2normProj2gamma}, we immediately get
    \begin{align*}
        h(\ab) 
        &= \lambda_2 \ab^\top \Qb^\top \Qb \ab + (\lambda_3-\lambda_2) \ab^\top \left(\Qb_3^\top \Qb_3+\Qb_4^\top \Qb_4\right) \ab \\
        &= \lambda_2 \|\gammab_1\|_2^2 + \lambda_2 \|\gammab_2\|_2^2 + \lambda_3 \|\gammab_3\|_2^2 + \lambda_3 \|\gammab_4\|_2^2,
    \end{align*}
    where the regularization on $\gammab_1$ and $\gammab_4$ are irrelevant as per the considerations in Proposition~\ref{prop:DDPCequivalence1}.  
\end{proof}
This regularization strategy can therefore be interpreted as aiming for the ``best of both worlds'' previously alluded to at the end of Section~\ref{sec:standard2Norm}. Essentially, instead of fully disregarding the input regularization term \eqref{eq:2NormCostTerm2}  or giving it the same weight as the output regularization term \eqref{eq:2NormCostTerm1}, both terms can be tuned individually. Similarly, the following regularization strategy combines the input regularization \eqref{eq:2NormCostTerm2} with an SPC predictor.

 \begin{prop}\label{prop:DDPCstrategy2}
     \!$\gammab$-DDPC with regularization $\tilde h(\gammab) = \!\lambda_a \|\gammab_2\|_2^2$ and constraint $\gammab_3 = \zerob$ is equivalent to DeePC with regularization $h(\ab) = \lambda_a \|\Pib\ab\|_2^2$ (or $h(\ab) = \lambda_a \|\ab\|_2^2$) and constraint $\Pib_\perp \ab = \zerob$. Furthermore, both schemes are equivalent to SPC with the additional regularization $h^\ast(\xib, \ubs_f, \ybs_f) = \lambda_a \|\ubs_f-\Ub_f \Wb_p^+\xib\|_{\Rbc_\text{reg}}^2$.
 \end{prop}
\begin{proof}
    For DeePC, it was already noted in \cite{Dorfler2021} that $\Pib_\perp \ab = \zerob$ is equivalent to enforcing the SPC constraint.
    With respect to $\gammab$-DDPC, note that every pair $(\xib, \ubs_f)$ is uniquely associated with a pair $(\gammab_1, \gammab_2)$ via 
    $$
        \begin{pmatrix}
            \gammab_1 \\ \gammab_2
        \end{pmatrix}
        = \Lb_{1:2}^{-1}
        \begin{pmatrix}
            \xib \\ \ubs_f
        \end{pmatrix}. 
    $$
    For $\gammab_3 = \zerob$, it also follows that 
    $$
        \ybs_f = \begin{pmatrix}
            \Lb_{31} & \Lb_{32}
        \end{pmatrix}
        \begin{pmatrix}
            \gammab_1 \\ \gammab_2
        \end{pmatrix} = \begin{pmatrix}
            \Lb_{31} & \Lb_{32}
        \end{pmatrix} 
        \Lb_{1:2}^{-1}
        \begin{pmatrix}
            \xib \\ \ubs_f
        \end{pmatrix}, 
    $$
    which is equivalent to constraining the output predictions to $\ybs_f\! =\! \hat\ybs_\text{SPC}(\xib, \ubs_f)$, since $\Kb_\text{SPC}\!
        =\! \Yb_f\! \Zb^+\! =\! \begin{pmatrix}
            \Lb_{31} & \Lb_{32}
        \end{pmatrix} \!
        \Lb_{1:2}^{-1}$.
        Furthermore, note that \eqref{eq:2NormCostTerm2} and \eqref{eq:2NormCostTerm3} originate from decomposing the following block-expression, which can be reformulated in terms of $(\gammab_1, \gammab_2)$ as
    $$
    \left\|\begin{pmatrix}
        \xib \\
        \ubs_f
    \end{pmatrix}\right\|_{\left(
    \Zb \Zb^\top\right)^{-1}}^2 \!\!= \! \left\|
        \Zb \ab
\right\|_{\left(
    \Zb \Zb^\top\right)^{-1}}^2 \!\!=  \! \|\Pi\ab\|_2^2 \!
     = \! \|\gammab_1\|_2^2 + \|\gammab_2\|_2^2. 
    $$
    Regarding the DeePC regularization, the choice between $h(\ab) = \lambda_a \|\Pi\ab\|_2^2$ or $h(\ab) = \lambda_a \|\ab\|_2^2$ is irrelevant, since $\ab = \Pib \ab + \Pib_\perp \ab=\Pib \ab $ due to the constraint $\Pib_\perp \ab=\zerob$.
    Regarding the $\gammab$-DDPC regularization, the result already follows from arguing that regularization of $\gammab_1$ is irrelevant (as in Prop.~\ref{prop:DDPCstrategy1}). 
    Yet, we even have a stronger connection here. To see this, note that \eqref{eq:2NormCostTerm3} can also be expressed via 
    $$
        \|\xib\|_{\left(\Wb_p\Wb_p^\top\right)^{-1}}^2 = \ab^\top \Wb_p^+ \Wb_p \ab = \ab^\top \Qb_1^\top \Qb_1 \ab = \|\gammab_1\|_2^2. \vspace{-0.5mm}
    $$
    Hence, we have $\tilde h(\gammab)= h^\ast(\xib, \ubs_f, \ybs_f)$ due to
    \begin{alignat*}{3}
        &\|\gammab_1\|_2^2 + \|\gammab_2\|_2^2  &&= \|\ubs_f-\Ub_f \Wb_p^+\xib\|_{\Rbc_\text{reg}}^2 + \|\xib\|_{\left(\Wb_p\Wb_p^\top\right)^{-1}}^2  \\
        &\iff \;\;\;\|\gammab_2\|_2^2& &= \|\ubs_f-\Ub_f \Wb_p^+\xib\|_{\Rbc_\text{reg}}^2. 
    \end{alignat*}
    \end{proof}
    This regularization strategy is combining the cautious input regularization with strict adherence to the SPC predictor. Note that for $\lambda_2 = \lambda_a$ and $\lambda_3 \to \infty$, the regularization strategy in Proposition~\ref{prop:DDPCstrategy1} tends towards the one in Proposition~\ref{prop:DDPCstrategy2}. However, we want to emphasize that ``tending towards'' does not mean equivalence. For example, there is still a significant difference between the two in the presence of output constraints $\ybs_f \in \Yc$, as observed in \cite{KLAEDTKE2023}. From the proofs in this section, we can also summarize the following relations between regularizing the $\gammab$-variables and their counterparts in terms of $\left(\xib, \ubs_f, \ybs_f\right)$. Note the resemblance to the characterization of DeePC regularization in \eqref{eq:2NormCost}. 
    \begin{cor}\label{corr:gammaDDPCcost}
        Regularization of the $\gammab$-variables can be equivalently expressed as
        \begin{align*}
            \|\gammab_1\|_2^2 &= \|\xib\|_{\left(\Wb_p\Wb_p^\top\right)^{-1}}^2,\\
            \|\gammab_2\|_2^2 &= \|\ubs_f-\Ub_f \Wb_p^+\xib\|_{\Rbc_\text{reg}}^2,\\
            \|\gammab_3\|_2^2 &= \|\ybs_f-\hat\ybs_\text{SPC}(\xib, \ubs_f)\|_{\Qbc_\text{reg}}^2.
        \end{align*}
        Further, recall that $\gammab_4$ characterizes the null space of the data matrices and does not affect $\left(\xib, \ubs_f, \ybs_f\right)$. Hence, $\|\gammab_4\|_2^2$ has no counterpart in terms of $\left(\xib, \ubs_f, \ybs_f\right)$.
    \end{cor}
    The Propositions~\ref{prop:DDPCequivalence1}-\ref{prop:DDPCstrategy2} and Corollary~\ref{corr:gammaDDPCcost} allow transferring virtually any results between the two frameworks, e.g., the main result in \cite[Thm.~1]{Breschi2023regularizationImpact} characterizing the effect of $\|\gammab_2\|_2^2$ for the input data trajectory being white noise. To summarize, while useful results have been derived within the $\gammab$-DDPC framework, these results do not depend on this specific framework. Furthermore, the decrease in number of optimization variables from $\dim(\ubs_f)+\dim(\ybs_f)+\dim(\ab)=m N_f + p N_f + \ell$ to $\dim(\ubs_f)+\dim(\ybs_f)+\dim(\gammab_2)+\dim(\gammab_3)=m N_f + p N_f + m N_f + p N_f$ is sometimes stated as a major advantage of this framework \cite{Breschi2022new}. However, this is just a generic preprocessing step associated with eliminating degrees of freedom represented by $\xib$ (via fixing $\gammab_1$) and by the null space of the data matrix (via dropping $\Qb_4$, respectively $\gammab_4$). For problem instances, where such a preprocessing step is tractable, and the dimensionality reduction is considered useful, one might as well forego the $\gammab$-DDPC framework by eliminating $\gammab_2$ and $\gammab_3$ via the corresponding cost in Corollary~\ref{corr:gammaDDPCcost}. However, this consideration is not unique to $\gammab$-DDPC, but also arises in the original DeePC framework, where $\ab$ can be eliminated using Lemma~\ref{lem:priceTag}.
\section{Conclusion and Outlook}\label{sec:conclusions}

In this note, we have established equivalences between the quadratic regularization strategies proposed in the DeePC and the $\gammab$-DDPC frameworks. 
Rather than claiming that one framework is superior to the other, these equivalences should be seen as an opportunity to transfer results between them. In future work, we will continue our efforts to unify DPC frameworks such as the one proposed in \cite{Lazar2023GeneralizedDP} by analyzing equivalences in terms of closed-loop behavior and predictive behavior (in the sense of implicit predictors, recently proposed in \cite{KLAEDTKE2023}).

\newpage
\footnotesize


\end{document}